\documentclass{PoS}

\title{P- and D-wave spin-orbit splittings in heavy-light mesons}

\ShortTitle{P- and D-wave spin-orbit splittings in heavy-light mesons}

\author{\speaker{Jonna Koponen}\\
        University of Helsinki\\
        E-mail: \email{jonna.koponen@helsinki.fi}}

\abstract{This is a summary of a detailed study of heavy-light meson excited state energies.
Our lattice measurements include both radial and orbital excitations. Particular attention
is paid to the spin-orbit splittings, to see which one of the states (for a given angular
momentum L) has the lower energy. In nature the closest equivalent of this heavy-light system
is the $B_s$ meson, which allows us to compare our lattice calculations to experimental
results (where available) or give a prediction where the excited states, particularly P-wave
states, should lie.}

\FullConference{The XXV International Symposium on Lattice Field Theory\\
		 July 30 - August 4 2007\\
		 Regensburg, Germany}

\usepackage{amsmath}

\begin{document}

\section{Introduction}

The excited state spectrum of $D_s$ and $B_s$ mesons has attracted a lot of interest lately.
For example, last year (2006) CDF and D\O~collaborations reported measurements of two $B_s$
meson P-wave states~\cite{CDF_D0}, and BaBar and Belle measured excited $D_s$ meson
states~\cite{Aubert:2006mh, Abe:2006xm}. Lattice QCD has now an excellent opportunity
to offer some knowledge in the matter from the theory side.
We have done an in-depth study of a heavy-light meson excited state energy spectrum on a
lattice. This proceedings paper offers a summary of our results. Full details of the study
can be found in~\cite{JK}.

We have measured the energies of both angular and first radial excitations of
heavy-light mesons. Since the heavy quark is static, its spin does not play a role
in the measurements. We may thus label the states as $\mathrm{L}_{\pm}=\mathrm{L}\pm\frac{1}{2}$,
where L is the orbital angular momentum and $\pm\frac{1}{2}$ is the spin of the light quark.

Our main measurements are done on a $16^3\!\times\! 32$ lattice with 160 configurations.
The two degenerate quark flavours have a mass that is close
to the strange quark mass (about 1.1$m_s$). The lattice configurations were generated
by the UKQCD Collaboration using lattice action parameters \mbox{$\beta = 5.2$},
$c_\textrm{SW} = 2.0171$ and $\kappa = 0.1350$. The lattice spacing is $\approx 0.11$~fm.
More details of the lattice configurations used in this study can be found in
Refs.~\cite{Allton,PRD69}. Because the light quarks are heavier than true \textit{u}
and \textit{d} quarks, the pion mass is $m_{\pi}=0.73(2)$~GeV. Two different levels of
fuzzing (2 and 8 iterations of conventional fuzzing) are used in the spatial directions
to permit a cleaner extraction of the excited states.

We also introduce two types of smearing in the time direction. First we try
APE type smearing, where the original links in the time direction are replaced
by a sum over the six staples that extend one lattice spacing in the spatial directions
(``sum6'' for short). To smear the static quark even more we then use hypercubic
blocking (``hyp'' for short), but again only in the time direction. The label ``static''
is used to denote the heavy quark that is not smeared in the time direction.

\section{Energy spectrum}

To obtain the energy spectrum we measure the 2-point correlation function
\begin{equation}
\label{2point}
C_2(T)=\langle P_t\Gamma G_q(\mathbf{x},t+T,t)P_{t+T}
\Gamma^{\dag}U^Q(\mathbf{x},t,t+T)\rangle \  ,
\end{equation}
where $U^Q(\mathbf{x},t,t+T)$ is the heavy (infinite mass)-quark propagator
and $G_q(\mathbf{x},t+T,t)$ the light anti-quark propagator. $P_t$
is a linear combination of products of gauge links at time $t$
along paths $P$ and $\Gamma$ defines the spin structure of the operator.
The $\langle ...\rangle$ denotes the average over the whole lattice.
A detailed discussion of lattice operators for orbitally excited mesons
can be found in~\cite{Lacock}. In this study, the same operators are used
as in~\cite{MVR}.
The energies ($m_i$) and amplitudes ($a_i$) are extracted by fitting the $C_2$
with a sum of exponentials,
\begin{equation}
\label{C2fit}
[C_2(T)]_{f_1,f_2}\approx
\sum_{i=1}^{N_{\textrm{max}}}a_{i, f_1}\mathrm{e}^{-m_i T}a_{i, f_2},\;
\textrm{where $N_{\textrm{max}}=2\textrm{ -- }4$, $T\leq 14$}.
\end{equation}
In most of the cases 3 exponentials are used to try to ensure the first
radially excited  states are not polluted by higher states.
Also 2 and 4 exponential fits were used to cross-check the results wherever
possible. Indices $f_1$ and $f_2$ denote the amount of fuzzing.

\begin{figure}
\centering
\includegraphics[height=0.60\textwidth, angle=-90]{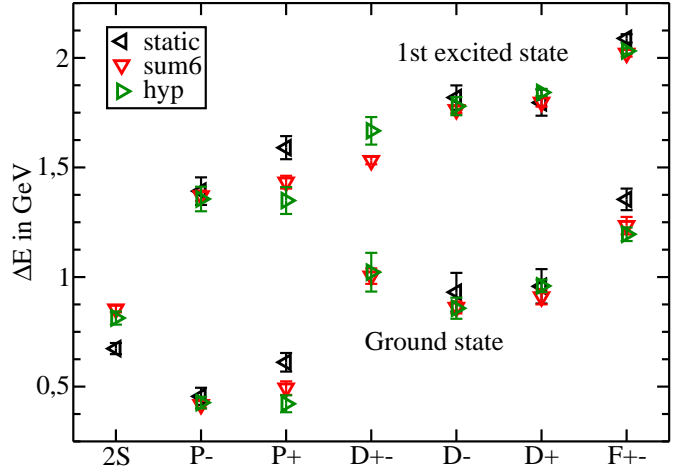}
\caption{Energy spectrum of the heavy-light meson.
Here L$+$($-$) means that the light quark spin couples to angular
momentum L giving the total $j=\textrm{L}\pm 1/2$. The 2S is the first radially
excited L$=0$ state. The D$+-$ is a mixture of the D$-$ and D$+$ states, and likewise
for the F$+-$.  Energies are given with respect to the S-wave
ground state (1S). Here $a=0.110(6)$~fm is used to convert the energies
to physical units. The error bars shown here contain statistical
errors only.}
\label{fig:espectr}
\end{figure}

\begin{table}
\centering
 \begin{tabular}{|c|ccc||c|ccc|}
 \hline
 nL$\pm$ &\, static\, &\, sum6\, &\, hyp\,  & nL$\pm$ &\, static\, &\, sum6\, &\, hyp\, \\
 \hline
 1S     & 3.60(4) & 2.805(14)& 2.51(2) &  2S     & 5.40(5) & 5.08(4)  & 4.67(8)  \\
 1P$-$  & 4.82(10)& 3.92(4)  & 3.64(7) &  2P$-$  & 7.31(16)& 6.45(6)  & 6.11(15) \\
 1P$+$  & 5.23(10)& 4.12(8)  & 3.63(10)&  2P$+$  & 7.83(13)& 6.62(8)  & 6.09(16) \\
 1D$\pm$&  -      & 5.48(10) & 5.22(17)&  2D$\pm$& -       & 6.88(4)  & 6.94(14) \\
 1D$-$  & 6.08(14)& 5.10(7)  & 4.79(12)&  2D$-$  & 8.44(10)& 7.50(4)  & 7.24(11) \\
 1D$+$  & 6.15(20)& 5.22(9)  & 5.06(7) &  2D$+$  & 8.38(15)& 7.59(5)  & 7.41(4)  \\
 1F$\pm$& 7.21(12)& 6.09(10) & 5.69(8) &  2F$\pm$& 9.16(3) & 8.18(4)  & 7.88(2)  \\
 \hline
\end{tabular}
\caption{Heavy-light meson energies on the lattice in units of $r_0$. 
Here $r_0/a=4.754(40)(+2-90)$ (Ref.~\cite{Allton}), which means $r_0\approx 0.52$~fm.
The n denotes the radial excitation and n$-1$ gives the number of nodes
in the wavefunction of the state.}
\label{EnergyTable}
\end{table}

The extracted energy spectrum is shown in Fig.~\ref{fig:espectr}. In most cases,
using different smearing for the heavy quark does not seem to change the energies
significantly --- the exceptions being the P$+$ (and excited D$+-$) state.
The energy of the D$+-$ state had been expected to be near the spin
average of the D$-$ and D$+$ energies, but it turns out to be a poor estimate of
this average. Therefore, it is not clear to what extent the F$+-$ energy
is near the spin average of the two F-wave states, as was originally hoped.

\subsection{Interpolation to the b-quark mass}

\begin{figure}
\centering
\includegraphics[angle=-90,width=0.95\textwidth]{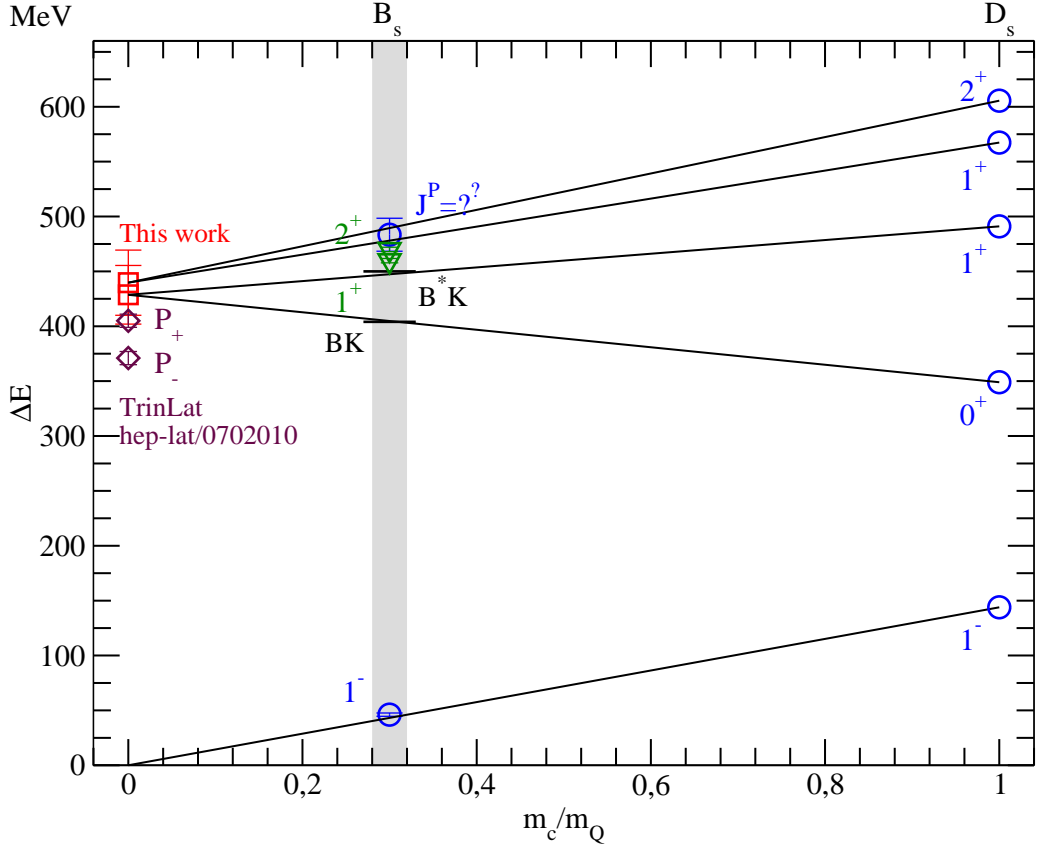}
\caption{Interpolation to the b-quark mass.
The ratio $m_c/m_b$ is taken to be $0.30(2)$ (from~\cite{PDG}; shown by the vertical
band). The $D_s$ meson experimental results are
from \protect\cite{PDG} (blue circles), and the $B_s$ meson experimental results are
from \protect\cite{PDG} (blue circles) and \protect\cite{CDF_D0} (green triangles).
TrinLat group's $m_c/m_Q=0$ results (maroon diamonds) are from~\protect\cite{II}.
Our results (using ``hyp'' configurations) are marked with red squares. The agreement
with the measured $B_s$ meson P-wave states is very good as seen in
Table~\protect\ref{BsEneTable}. If the ``sum6'' lattice results were used instead
of the ``hyp'' results, the predicted energies at $m_c/m_b$ would be too high compared
to experiments.}
\label{fig:Bs_interpolation}
\end{figure}

\begin{table}
\centering
 \begin{tabular}{|c||c|c|c|}
 \hline
 J$^P$ & hyp & sum6   & experiments \\
 \hline
 $0^+$ & $405 \pm 20$ MeV & $399 \pm 12$ MeV & -  \\
 $1^+$ & $447 \pm 20$ MeV & $441 \pm 12$ MeV & -  \\
 \hline			                     
 $1^+$ & $477 \pm 22$ MeV & $517 \pm 22$ MeV & $460 \pm 3$ MeV \\
 $2^+$ & $489 \pm 22$ MeV & $528 \pm 22$ MeV & $470 \pm 3$ MeV \\
 \hline
\end{tabular}
\caption{Our predictions for $B_s$ meson mass differences, M(B$^\ast_s$)-M(B$_s$),
for the P-wave states. The uncertainty in the ratio $m_c/m_b$ was not taken into
account in the error estimates. The experimental results are from \cite{CDF_D0, PDG}.
Here the ``hyp'' lattice P$+$ energy is taken to be E(P$-$) + spin-orbit splitting = 440(30) MeV
instead of the direct lattice measurement E($P+$) = 423(39) MeV. The latter would
lead to predictions E($1^+$) = 466(28) MeV and E($2^+$) = 477(28) MeV for the
$B_s$ meson.}
\label{BsEneTable}
\end{table}


\begin{figure}
\centering
\includegraphics[angle=-90,width=0.95\textwidth]{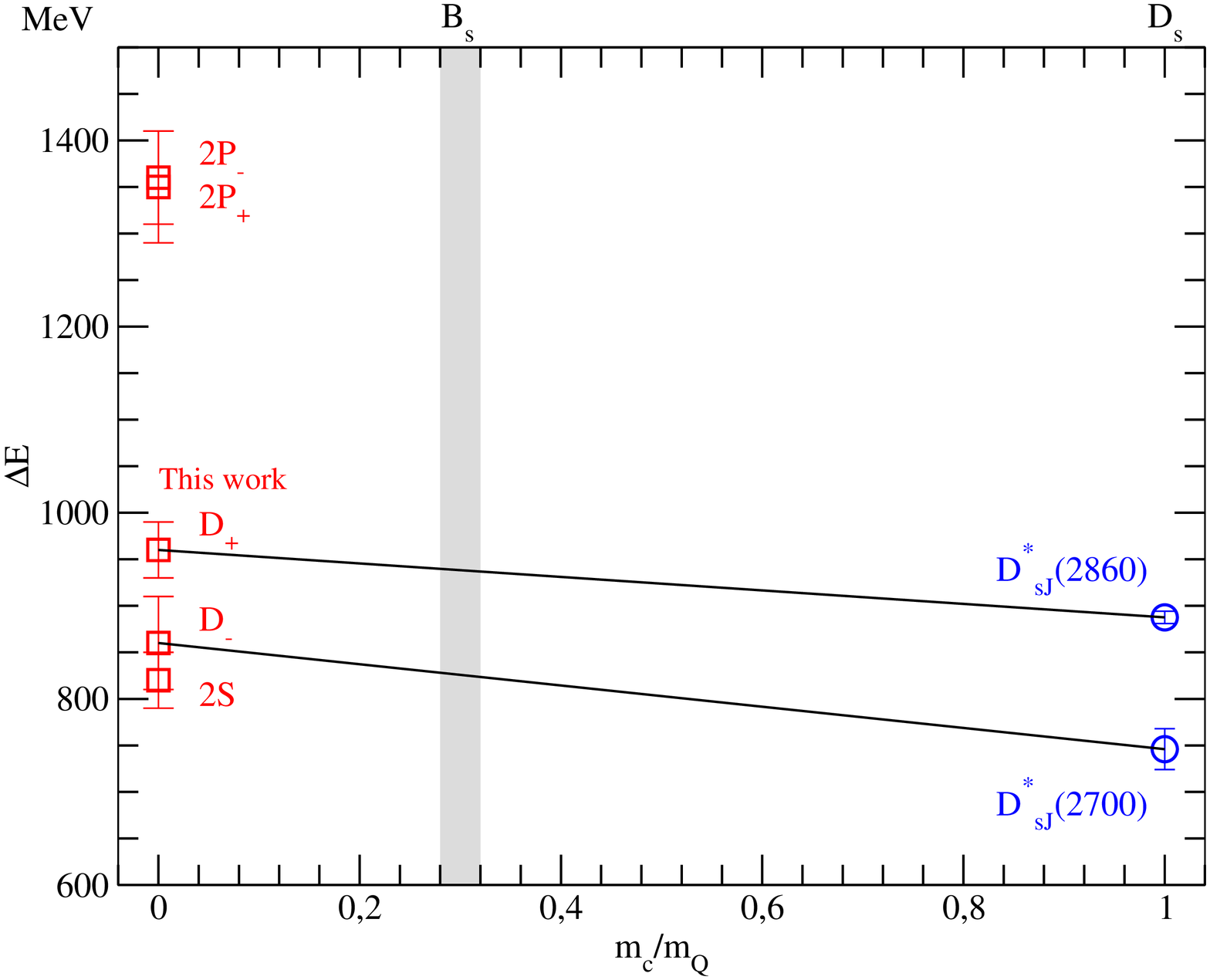}
\caption{Interpolation to the b-quark mass: higher
excited states. The lines illustrate what the interpolation would look
like, if the $D_s$ meson states were D-wave states. The experimental
results are from~\protect\cite{Aubert:2006mh, Abe:2006xm}.
Interpolating to $m_c/m_b$ predicts D-wave $J^P=3^-$, $J^P=1^-$
at 938(23) and 826(42) MeV respectively.}
\label{fig:Bs_interp2}
\end{figure}

One way to obtain predictions of the $B_s$ meson excited state energies is 
to interpolate in $1/m_Q$, where $m_Q$ is the heavy quark mass, between
the static heavy quark lattice calculations
and $D_s$ meson experimental results, i.e. interpolate
between the static quark ($m_Q=\infty$)  and the charm quark
($m_Q=m_c$). Here we, of course, have to assume that the measured
$D_s$ meson states are simple quark--anti-quark states. This is not
necessarily true: for example the mass of the $D_{s0}^\ast(2317)$ is
much lower than what is predicted by conventional potential models, and it has thus
been proposed that it could be either a four quark state, a $DK$ molecule or
a $D\pi$ atom. However, the inclusion of chiral radiative corrections could change
the potential model predictions considerably~\cite{LeeLee}. 
Here we assume simple quark--anti-quark states and do a linear interpolation
(see Fig.~\ref{fig:Bs_interpolation}). The two lowest P-wave states seem to lie
only a couple of MeV below the $BK$ and $B^\ast K$
thresholds respectively. Our predictions are given in Table~\ref{BsEneTable}.

As for other excited states, BaBar and Belle observed two new states,
$D_{sJ}^\ast(2860)$ and $D_{sJ}^\ast(2700)$, in 2006
\cite{Aubert:2006mh, Abe:2006xm}. The $J^P$ quantum numbers of the
$D_{sJ}^\ast(2860)$ can be $0^+$, $1^-$, $2^+$, etc., so it could
be a radial excitation of the $D_{s0}^\ast(2317)$ or a $J^P=3^-$
D-wave state. The first interpretation is rather popular, but our lattice
results favour the D-wave $J^P=3^-$ assignment in  agreement with
Colangelo, De Fazio and Nicotri \cite{Colangelo}.
Interpolation then predicts a D-wave $J^P=3^-$ $B_s$ state at 938(23) MeV.
The alternative interpretation as a 2P state, when compared with our lattice result,
would lead to an unreasonably large $1/m_Q$ dependence.
In addition, the $D_{sJ}^\ast(2700)$ could be a radially excited S-wave state or a
D-wave $J^P=1^-$ state. If the latter identification is assumed,
then a D-wave $J^P=1^-$ $B_s $ state at 826(42) MeV is expected (see
Fig.~\ref{fig:Bs_interp2}).

\subsection{Spin-orbit splitting}

\begin{figure}
\centering
\includegraphics[height=0.65\textwidth, angle=-90]{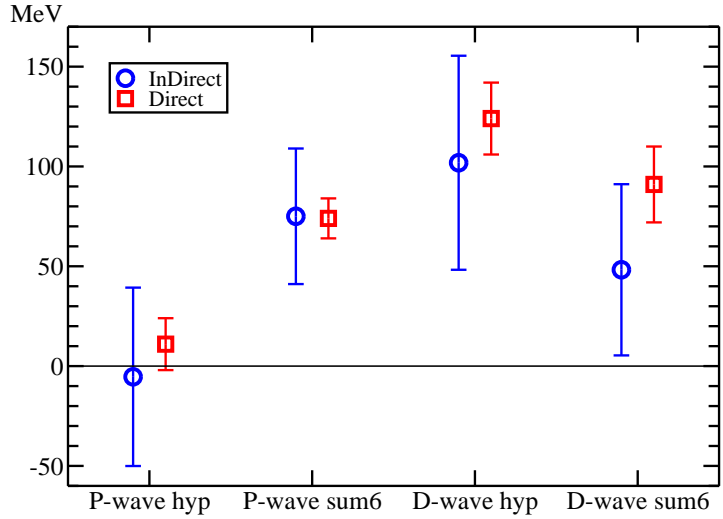}
\caption{The spin-orbit splittings of P-wave and D-wave states. The P-wave
spin-orbit splitting is small (consistent with zero), whereas the D-wave spin-orbit
splitting is much larger and positive.}
\label{fig:SOS}
\end{figure}

\begin{table}
\centering
 \begin{tabular}{|c|c|c|}
 \hline
 E(nL$+$) $-$ nE(L$-$) & Indirect     & Direct \\
 \hline
 E(1P$+$) $-$ E(1P$-$) & $-5 \pm 45$~MeV   & $11 \pm 13$~MeV  \\
 E(1D$+$) $-$ E(1D$-$) & $102 \pm 54$~MeV  & $124 \pm 18$~MeV \\
 E(2P$+$) $-$ E(2P$-$) & $-10 \pm 83$~MeV  & - \\
 E(2D$+$) $-$ E(2D$-$) & $70 \pm 60$~MeV   & - \\
 \hline
\end{tabular}
\caption{P-wave and D-wave spin-orbit splittings using ``hyp'' configurations.}
\label{SOSTable}
\end{table}

We pay particular attention to the spin-orbit splitting (SOS) of the P-wave states,
i.e. the energy difference of the 1P$+$ and 1P$-$ states. We extract the SOS in
two different ways:
\begin{enumerate}
\item
\emph{Indirectly} by simply calculating the difference using the energies given by
Eq.~\ref{C2fit}, when the P$+$ and P$-$ data are fitted separately.

\item
Combining the P$+$ and P$-$ data and fitting the ratio
$\textrm{C}_2(\textrm{P}+)/\textrm{C}_2(\textrm{P}-)$, which enables us
to go \emph{directly} for the spin-orbit splitting, $m_\textrm{1P$+$}-m_\textrm{1P$-$}$.
\end{enumerate}
The D-wave spin-orbit splitting is also extracted in a similar manner.
The results of the fits are given in Table~\ref{SOSTable} and in Fig.~\ref{fig:SOS}.
The two smearings, ``sum6'' and ``hyp'', give different spin-orbit
splittings for the P-wave: 74(10) MeV and  11(13)~MeV, respectively.
This is unexpected, and is studied in detail in~\cite{JK}. A similar
difference is not seen in the D-wave, where direct analysis gives 124(18)~MeV (``hyp'')
and 91(19)~MeV (``sum6'').

\section{Conclusions}

\begin{itemize}
\item
With the ``hyp'' lattice, our predictions for the $1^+$ and $2^+$ P-wave state masses agree
very well with the experimental results. We also predict that the masses of the two
lower P-wave states ($0^+$ and $1^+$) should lie only a few of MeV below the $BK$
and $B^\ast K$ thresholds respectively.

\item
Also with the ``hyp'' lattice, the P-wave spin-orbit splitting is small
(essentially zero), but the D-wave spin-orbit splitting is clearly non-zero
and positive. In contrast, another lattice group finds the D-wave spin-orbit splitting
to be slightly negative (see~\cite{II}), i.e. they seem to observe the famous inversion
\cite{Schnitzer}. On the other hand, in~\cite{LeeLee} Woo lee and Lee suggest that
the absence of spin-orbit inversions can be explained by chiral radiative corrections
in the potential model.

\end{itemize}

\section*{Acknowledgements}

I am grateful to my  collaborators, Professors A.M. Green and  C. Michael, and to
the UKQCD Collaboration for providing the lattice configurations.
I wish to thank Professor Philippe de Forcrand for useful comments
and also the Center for Scientific Computing in Espoo,
Finland, for making available the computer resources.
This work was supported in part by the EU Contract No. MRTN-CT-2006-035482,
``FLAVIAnet''.

\end{document}